# Intensity-modulated optical fiber strain sensor for continuous measurements of below-the-surface food deformation during drying

Hamed Jafarishad [a], Mucheng Li [a], Yao Shen [a], Pawan Singh Takhar [b], Yuxiang Liu [a, *]

[a] Department of Mechanical Engineering, Worcester Polytechnic Institute, Worcester, Massachusetts, 01609, USA
[b] Department of Food Science and Human Nutrition, University of Illinois at Urbana-Champaign, Champaign, Illinois, 61801, USA

**Abstract**

We present a mm-scale optical fiber strain sensor to continuously measure the food deformation below the surface during the drying process. The sensor has a soft buffer sleeve, which enables a large measurable deformation range and minimizes the influence on the native food deformation. The sensor performance was investigated both theoretically and experimentally. To demonstrate the application in foods, we embedded the sensor inside fresh banana slices during a 4-hour-long air-drying process, where its wet basis moisture contents dropped from around 80% to 40%. The fiber sensor measurements covered the full normal strain of ~20% of the banana slices and revealed the shell-hardening characteristics in both the spatial and time dependences, by the comparison with the surface strains measured by the computer vision. Our work provides an unprecedented strain sensor that holds a high potential to contribute to improving both fundamental understandings and process monitoring of food drying processes.

**Keywords:** Optical fibers, Strain sensors, Drying, Food deformation

## 1- Introduction

Foods deform when dried. [1] The deformation is induced by the moisture loss from and movement in the food matrix, among other complex physical and chemical changes. Food deformation is related to food quality parameters such as the texture, moisture content, crust formation, wholesomeness, and crack formation [2], all of which evolve during the whole drying process [3]. Knowledge of such continuously changing food deformation is important both to enhance the fundamental understanding and to improve the control of the drying processes. However, the intrinsic characteristics of foods and the process environments vary vastly. It remains a challenge, even for some common foods, to measure the deformation during drying, despite of the strong need in food science and engineering. Some of the challenges are listed in the following. First, the deformation range of food can be large, up to 30% to 70% [4], during the whole drying process. Knowledge of the deformation is especially important at the end of the drying process, when the deformation is the largest and the measurements are most difficult, for making a decision to stop the drying. Further, food deformation is susceptible to changes induced by the insertion of sensors that are stiffer than the foods. Most food samples have nonuniform and anisotropic food matrices while others have high moisture contents before drying, so it is difficult to create flat and dry surfaces on these food samples in order to implement conventional thin-film strain gauges. Furthermore, the drying ovens and the harsh process environments often introduce additional concerns on the compatibility of the sensors. Due to the difficulties in experimental measurements, there have been mathematical models developed to evaluate food deformation theoretically [1, 5]. These models have been applied to several types of foods, such as corn kernels [6, 7], strawberries and carrots [8], apples [9], and potatoes [10]. However, it is difficult for these models to account for all the details of the real-world food matrix, such as material inhomogeneity and anisotropy. Hence, experimental deformation measurements are beneficial, if not necessary, to verify the faithfulness of these models.

[*] Corresponding author.
*E-mail address:* yliu11@wpi.edu (Y. Liu).

Existing experimental tools for deformation measurements can be categorized into contact-based and non-contact-based methods, while not all of them are suitable for food samples. The most commonly used method is the thin-film resistive strain gauge [11], often simply called strain gauge, which is a contact-based method. The strain gauge is composed of a conductive wire that is embedded in a plastic film in a zig-zag pattern of parallel lines. When the strain gauge is glued to the surface of a sample, the sample deformation changes the conductive wire's length and hence its electrical resistance, the latter of which is measured to back out the strain. Alternatively, piezoelectric strain sensors with a similar form factor have also been reported [12]. However, strain gauges or piezoelectric strain sensors have not been applied to foods, to the best of our knowledge, with the reasons specified in the following. Depending on the food, these strain gauges/sensors can be stiffer than the sample and hence cannot provide faithful measurements. Their electrical readouts are susceptible to the interference of electromagnetic (EM) waves existing in the environment [13] and cannot be used near strong electrical currents or in microwave dryers. The implementation of these gauges/sensors requires a flat, smooth, and hard surface on the sample to ensure a good adhesion, while such a surface is rarely available in foods. The thin-film strain gauges/sensors are generally centimeter-long, too large for small food samples such as grains. Although not commonly used, mechanical loadings [14] have been used [15, 16] to validate mathematical models of food deformation and mechanical properties. However, the loads are applied over a large area of food samples, resulting in average measurements over a large area. In addition, such loadings must be applied by external equipment, difficult for either in-situ or continuous strain measurements in drying. Different from the abovementioned contact-based methods, non-contact methods such as cameras and image processing techniques [9] have been used for food samples, such as bananas [17], apples [18], pears [19], and potatoes [20, 21]. Although these imaging-based non-contact methods do not suffer from the adhesion and EM challenges, they can only measure strains on the exposed surfaces of foods. It is often challenging to install imaging devices inside a hot oven where the high temperature can endanger the electronics, while cameras installed outside have compromised quality and distortion of images due to the air currents and particles in the light path. In summary, current electrical strain sensors cannot satisfy the need of continuous, in-line strain measurements in foods during drying.

Alternative to electrical sensors, optical fiber sensors have great potential to enable the food strain measurements in harsh environments. Optical fibers, thanks for the low cost and simplicity of the structure, have been utilized for the strain sensor development [12]. A typical working principle of optical fiber strain sensors is to convert the sample deformation into the change of a sensing element on or next to the fiber, the latter of which is often a length or a position change and can be directly detected through the fiber. In this sense, most fiber strain sensors are essentially displacement sensors. Intensity-modulated strain sensors [22, 23], in which the readout is the optical intensity and is modulated by the external strain, are one of the most common fiber strain sensors [23]. Various applications for the optical fiber strain sensors have been reported, including structural health monitoring of buildings and bridges [24, 25], vehicle fatigue testing [26], and damage detection [27]. Despite their wide applications in other engineering areas, optical fiber strain sensors have not yet been applied in food measurements, not to mention in drying processes, to the best of our knowledge. The exiting designs of fiber strain sensors cannot address the challenges posed by soft food materials during drying. For example, the abovementioned optical fiber strain sensors are much stiffer than soft foods, such as fruits and vegetables [23]. Existing fiber strain sensors have limited measurable strain range that cannot cover the large deformation typically occurring in foods. For example, all the above fiber sensors can only measure strains up to 1%, far below the common strain range in foods. As a result, the current sensors cannot cover the large range of strains in foods. There is a need for a fiber optical strain sensor capable of in-line, continuous measurements of large strains in food during drying. Ideally, such a sensor should be small, soft, readily mountable inside food samples, and compatible with harsh environments such as strong EM waves.

In this study, we present an intensity-modulated optical fiber strain sensor specifically designed for soft food measurements during drying. This sensor can measure the strain inside food samples by measuring the optical intensity at the readout. It is immune to EM waves in drying chambers because there is no electrical conductor or signal in the sensor and because the frequency of the optical signal (>100 THz) is much higher that of environmental EM waves (~GHz). The intensity-modulated sensor mechanism has been chosen because of the simple design and fabrication process. With a small footprint and soft shell, this sensor can be applied to a large variety of food samples with different sizes and softness. In addition to systematic investigation of the sensor performance, we demonstrated the sensor capability of continuous, in-situ strain measurements in fresh banana slices over an air-drying process. Thanks to the unique design feature of a buffer sleeve, the sensor can measure large food deformation with a length change up to 20%, which is enough to cover the deformation range of considerable kinds of foods, such as fresh bananas, during drying. Additionally, it has a potential of reaching an even larger range by slight design modifications. Furthermore, the sensor is designed to enhance the attachment to food samples without the need of smooth or flat surfaces. This sensor is not limited to foods, but can be potentially employed to measure the large deformation of other challenging soft materials, such as pulp and paper, soil, and muscular and neural tissues.

## 2- Design and working principles

Before we proceed with the description of our work, we would like to clarify the terminology of "strain," which will be used extensively in the rest of the paper. In the conventional engineering context, mechanical deformation of an object is quantified by strains, where normal strains and shear strains define the changes of size and shape, respectively. Particularly, a normal strain is defined as the ratio of the length change to the initial length between two points that are fixed to and move with the object during its deformation. A positive normal strain corresponds to the elongation and negative to compression. In this paper, we aim to develop our sensors to measure the normal strain along the sensor direction, which can be any desired direction in the food, and we will not measure any shear strains. Therefore, the strain in the rest of the paper refers to the normal strain to avoid redundancy, unless specified. Moreover, we are only interested in measuring the shrinkage of the food samples, where the normal strain should always be negative. Therefore, we will describe the normal strain by its absolute value without bringing the negative sign. For example, we aim to measure a compressive food strain of 20%, without using -20% in the text.

The designed optical fiber strain sensor consists of a buffer sleeve, glass tube, metal disk, and optical fiber, all of which are depicted in Fig. 1(a). All the optical fibers in this work are single-mode (SMF-28, Corning) and the used optical wavelength is around 1300 nm. The fiber is inserted from the left end of the tube. At the right end, the glass tube is glued to a metal disk, serving as a mirror. When light is emitted from the fiber tip, the reflected light from the metal disk is collected by the same fiber, with the collected light power dependent on the fiber-disk distance due to the light divergence. The buffer sleeve is made of silicone softer than skin and is glued to the outer surface of glass tube and the optical fiber portion outside the tube. When mounted inside a food sample, the outer surface of the buffer sleeve is in direct physical contact and deform together with the food. When the sample deforms during drying, the deformation of the buffer sleeve leads to the slide of the fiber inside the glass tube (Fig. 1(b)), resulting in the fiber-disk distance change and hence the collected light power change, the monitoring of the latter of which allows the measurements of the food strain along the fiber direction. A schematic of the sensor readout system is illustrated in Fig. 1(c). Light from a pigtailed benchtop super-luminescent diode (SLD, S5FC1021S, Thorlabs) with a center wavelength and bandwidth of 1310 nm and 85 nm, respectively, goes through the fiber circulator (CIR-3-1290-1620-L-10-FA-5.5x38, Ascentta) and emitted from the fiber tip in the sensor. The collected portion of the reflected light travels backward along the fiber, through the fiber circulator, and its power is measured by the photodetector (PDA20C, Thorlabs). The output of the photodetector is a voltage proportional to the received optical power.

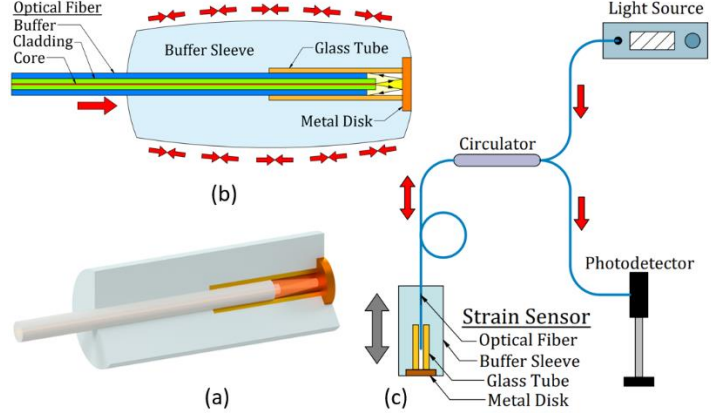

Fig. 1. Design of optical fiber strain sensor. a) 3D model and b) 2D schematic of the optical fiber strain sensor. c) Sensor readout system. All the blue lines are optical fibers.

One of the main features in our design that differ from the other fiber strain sensors is the buffer sleeve. The functions of this buffer sleeve have four folds. 1) It holds the fiber position inside the tube and prevents the fiber from slipping freely; 2) When the food deforms, the buffer sleeve moves the fiber inside the tube, changes the fiber-mirror distance, and modulates the optical intensity received by the optical fiber, allowing for reliable measurements of the food deformation. 3) The buffer sleeve is softer than human skin and small (~1.2 mm in diameter) in size, minimizing the influence on the food deformation when the food has similar or higher stiffness. 4) The buffering effect allows smaller strain inside the sleeve than the outside food, reducing the displacement of the fiber, which enables a large strain measurement range. To quantify the buffering effect, we define the buffering ratio as

$$b_r = \frac{\Delta L}{\Delta \delta}, \tag{1}$$

where $L$ is the length of the food sample in contact with the outer surface of the sensor, $\delta$ is the fiber-disk distance, and $\Delta L$ and $\Delta \delta$ are the changes of $L$ and $\delta$, respectively. When the thickness of the buffer sleeve approaches, $\Delta L$ approaches $\Delta \delta$ and $b_r$ approaches 1, indicating no buffering effect. The larger $b_r$, the higher the buffering effect, and the smaller the fiber-disk distance change compared with the food deformation.

In order to understand the role of the buffer sleeve and hence obtain a proper design, we used a finite element method (FEM) with commercial software (COMSOL Multiphysics, COMSOL Inc.) to simulate the sensor mechanical response to the deformed food on the outer sidewall. 2D simulation was performed on a single cross-section passing the axis of the sensor, with an axisymmetric model in the Structural Mechanics module in COMSOL. The geometrical and material properties of the sensor used in the simulation are listed in Table 1. The FEM meshed model is shown in Fig. 2(a). This model includes the optical fiber, glass tube, metal disk, and buffer sleeve, as shown in Fig. 2(b). We consider the deformation of the food during drying to be a slow process, so the elastic behavior of the buffer sleeve is assumed in the model. The glass tube, optical fiber, and metal disk are all assumed to be rigid because they are much stiffer than the buffer sleeve. No sliding is considered on the contact surfaces between the inner sidewall of buffer sleeve and the fiber, between the inner sidewall of buffer sleeve and glass tube, or between the outer sidewall of buffer sleeve and the food sample. We consider that the sensor is fully embedded inside the food sample and the physical contact between the food and sensor is on the cylindrical outer surface of the sleeve. By assuming a uniform food shrinkage on the contact surface, we apply a uniform shear strain on the outer cylindrical surface of the buffer sleeve in the direction of the sensor axis. The calculated sensor response is shown in Fig. 2(c), when there is a negative (compressive) 20% food strain applied on the outer surface of the buffer sleeve. A shrinkage of 0.6 mm occurs on the outer surface of buffer sleeve. By comparison, the change of the fiber-disk distance is about 0.2 mm, which, based on Eq. 1, results in a buffering ratio of ~3. The contour of von Mises stress, which is the stress equivalent to distortion, is presented in Fig. 2(d). The maximum stress in the buffer sleeve is about 200 kPa and occurs at three locations: the left and right ends of the sensor and the left end of glass tube, all three on the inner sidewall of the buffer sleeve. The high stress at the two ends of the sensor results from the large deformation mismatch between the outer and inner sidewall of the sleeve. The high stress at the left end of the glass tube indicates that the fiber slides into the tube, pinching the sleeve at this location, which might not occur in reality, thanks to the difficulty to realize a conformal adhesion between the sleeve and the fiber or tube at this location. Nevertheless, the safety of the sleeve under these stresses is not a concern because the silicone rubber (sleeve material, detailed in Section 3) has a tensile stress of 2.0 MPa. The contour of the normal strain along the axis of the sensor is shown in Fig. 2(e). The deformation is highest on the outer surface and keeps decreasing with the depths below it, except for the fiber entrance of the tube due to the abovementioned pinching effect.

| Geometrical Properties | | Material Properties | |
|---|---|---|---|
| Sensor length | 3.0 mm | Buffer sleeve, E | 0.5 MPa |
| Sensor diameter | 1.2 mm | Buffer sleeve, $\nu$ | 0.48 |
| Optical fiber diameter | 0.25 mm | Optical fiber | Rigid |
| Glass tube length | 1.5 mm | Glass tube | Rigid |
| Glass tube diameter | 0.36 mm | Metal disk | Rigid |
| Metal disk diameter | 0.6 mm | | |

Table 1. Geometrical and material properties of the designed sensor.

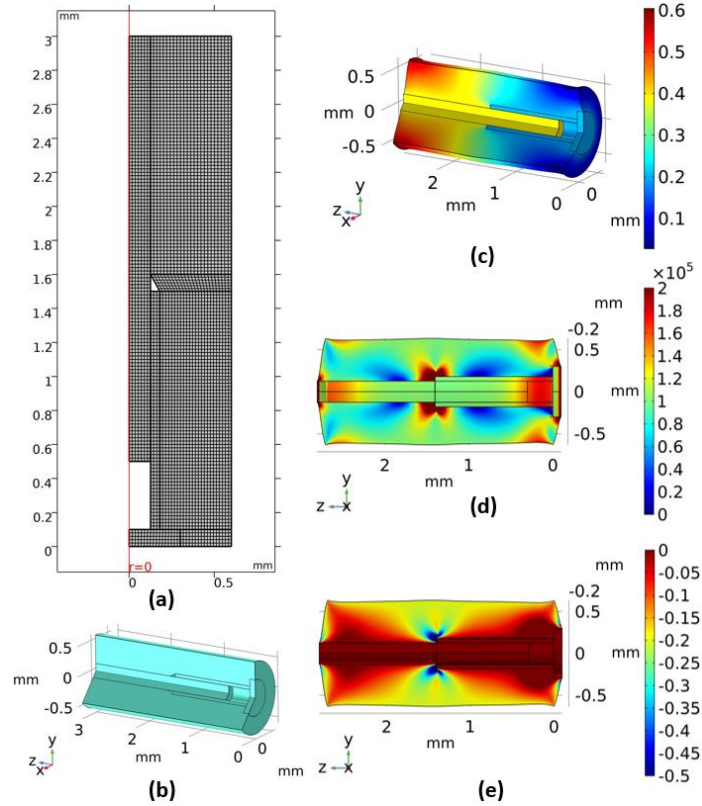

Fig. 2. a) Meshed model and b) 3D perspective view of the mechanical FEM model to simulate the sensor deformation. Simulation results of the distributions of the c) displacement, d) von Mises stress, and e) normal strain along the sensor axis, in response to a compressive 20% strain applied to the outer surface of the buffer sleeve.

These simulation results confirm the intended benefits of the buffering effect, which allows the measurements of high (20%) food strains with the fiber-tip distance change 3 times smaller than without the buffering layer. Such a scaledown is necessary because there is a higher boundary in the optically measurable fiber-disk distance. As the fiber tip goes too far away from the mirror, the reflected light is too weak to differentiate meaningful changes in distance. We can increase the thickness or the length of the buffer sleeve to obtain a higher buffering ratio and hence a higher measurable food strain range, at a cost of a larger sensor size. Although small, each strain sensor still occupies a finite volume in the food, so the measured strain essentially is averaged over the occupied volume. A larger sensor size will increase such an averaging effect, which is not desired considering the inhomogeneous nature of food deformation. Therefore, in order to determine a proper sleeve thickness in our sensor design, we need to further understand the measurable range of fiber-disk distance. In another word, if we would like to use the sensor design parameters in Table 1 to measure 20% compressive food strain, we should confirm 0.2 mm fiber-disk distance is practically measurable.

Such an experimental study was conducted with the setup shown in Fig. 3(a). A single-mode optical fiber (SMF-28, Corning) was fixed on a home-made gripper holder, which was in turn fixed on a translational motion stage. The optical fiber was inserted and could readily slide inside a glass tube (1068150026, Molex), which was fixed to the optical table with a copper disk glued at the tube bottom. The motion of the stage in the vertical direction was controlled by a servomotor (Z825B, Thorlabs) to change the fiber height and hence the fiber-disk distance. We used the same readout system as Fig. 1(c), with the sensor replaced by the setup in Fig. 3(a), to characterize the dependence of optical power readout on the fiber-disk distance ($\delta$). A schematic of the setup is illustrated in Fig. 3(b). As the servomotor moved the stage vertically, the distance between the fiber tip and copper disk changed which resulted in changes in the power of reflected light.

The obtained experimental results are plotted in Fig. 3(c). As the fiber tip got closer to the copper disk, the measured power increased exponentially. This increase results from the divergent nature of fiber-emitted light and can be well explained by the equations describing the free-space optical propagation, detailed in the following. Since we used single-mode optical fibers throughout this work, the light emitted from the fiber tip was always a fundamental Gaussian mode, which has the smallest beam waist at the tip surface and gradually increase the spot size with the propagation. After the light is emitted from the fiber, propagate to the disk, reflected, and travel back to the fiber, the optical power that is collected and detected by the fiber can be described as [28]

$$P_d = P_t \frac{1 - e^{-2(\omega/\omega_0)^{-2}}}{1 - e^{-2}}, \tag{2}$$

where $P_d$ and $P_t$ are the collected power and total emitted, respectively. $\omega_0$ and $\omega$ are the mode field radii of the light beam at the initial (the emission end face of the single-mode fiber) and light-collection (the same end face of fiber but after the light round trip) locations, respectively. In air, $\omega$ is a function of $\delta$ and can be calculated as

$$\omega(2\delta) = \omega_0 \sqrt{1 + \left(\frac{2\delta\lambda}{\pi\omega_0^2}\right)^2}, \tag{3}$$

where $\lambda$ is the wavelength of the light. With Eqs. 2 and 3 combined and a consideration of fiber tip internal reflection added, the dependence of detected power on $\delta$ can be written as

$$P_d = A\left(1 - e^{-2\left[1+\left(\frac{2\delta\lambda}{\pi\omega_0^2}\right)^2\right]^{-1}}\right) + B, \tag{4}$$

where $A$ is $P_t/(1-e^{-2})$. $B$ is not obtained from Eqs. 2 and 3, but a constant added to account for the internal reflection at the fiber tip, which results in a non-zero detected power in the experiment even if $\delta$ approaches infinity. In practice, $A$ and $B$ are parameters to be determined by fitting to experimental measurements. $\lambda$ and $\omega_0$ in our experiments are 1.31 µm and 4.6 µm, respectively. As shown in Fig. 3(c), our experimental measurements can be well fitted by the theoretical model described by Eq. 4 with a $R^2$ coefficient of determination of 0.987, and the fitting parameter values of *A* and *B* are 11.6 V and 0.659 V, respectively. Such a good fitting confirms the faithfulness of our theoretical model, supports our sensor design and our decisions on sensor parameters, and allows the conversion from the photodetector voltage readout to fiber-mirror distance change, the last of which will be used in the rest of the paper in both the sensor calibration and banana testing.

According to Fig. 3(c), there was a 460 µm dynamic range of $\delta$, i.e., between 40 and 500 µm, where $\delta$ can be reliably measured based on the reflected optical power. This dynamic range is more than twice larger than the ~200 µm of the fiber-disk distance calculated from our FEM model, which is needed to achieve the goal of measuring 20% food strains. In principle, we could decrease the buffer sleeve thickness or length further, resulting in a smaller sensor, without compromising the goal. However, we decide to use the design in Table 1 for the rest of the paper as a proof of concept. This leaves a safety factor of ~2.3 in the measurable strain range to cover any factors that cannot be well controlled in the sensor fabrication and implementation, and our fiber strain sensor should be able to measure food strains up to and possibly higher than 20%.

We would like to provide more details on how the $\delta$ range of 40~500 µm was determined. On the lower end, as the fiber tip went too close to the mirror ($\delta < 40$ µm), $\delta$ approached the coherent length (~20 µm) of our light source, resulting in an interference pattern and a sinusoidal dependence of the intensity on $\delta$ (data not shown). This dependence is not desired in the current intensity-modulated sensor design and should be avoided. Therefore, we did not record the data when $\delta < 40$ µm. On the higher end, there was no hard cutoff of the $\delta$ range, so we chose 500 µm with certain ambiguity. For example, we believe the distance could still be measured by the optical power at $\delta \sim 520$ µm, given the low power fluctuation at a fixed $\delta$. We note that the larger $\delta$ will reduce the sensitivity and resolution of the strain measurements, due to the smaller slope of power-$\delta$ curve in Fig. 3(c). As can be seen later in Sections 3~5, we targeted the initial $\delta$ value of 300 µm in our sensor fabrication (marked as a blue cross in Fig. 3(c)), where the $\delta$ range of 50~300 µm was needed to cover the 20% compressive food strains with a purpose to avoid a higher $\delta$ range and the corresponding low sensitivity and resolution, and to avoid a lower $\delta$ range close to coherent length. If one needs to measure elongation rather than compression of the food sample, the initial $\delta$ in the fabrication should be chosen to be closer to 40 µm. In this case, the elongated food sample moves the fiber away from the mirror during drying, and the same sensor design and the calibration curve in Fig. 3(c)) will be able to cover 20% or more positive normal strain.

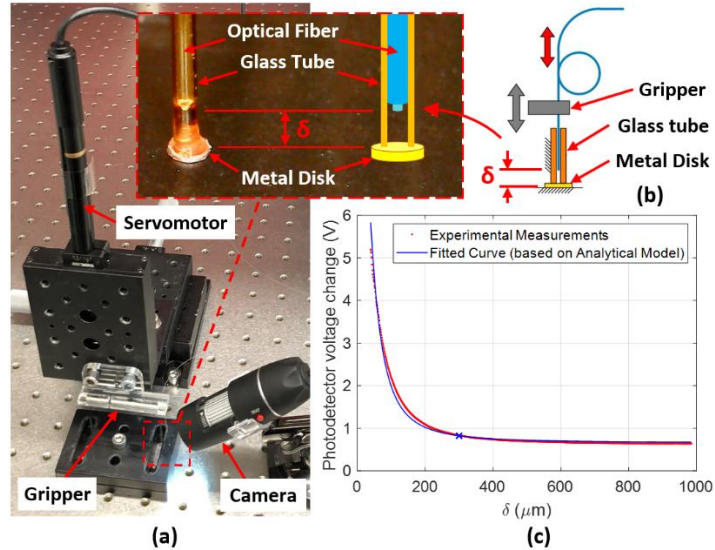

Fig. 3. Measurements of power of detected light versus the fiber-disk distance. a) Experimental setup. (Inset) closed-up of the fiber tip and disk. b) Schematic of the setup (the blue line is optical fiber). c) Dependence of detected power on the fiber-disk distance ($\delta$). The blue cross indicates the desired $\delta$ used in the fabrication process, which was also the initial value in the calibration and measurements.

## 3- Materials and fabrication

The fabrication process of the sensor can be completed in three phases, namely the fabrication of the buffer sleeve, fabrication of the reflective cap, and assembly. The process is detailed below in this section.

The fabrication started with Phase I, the fabrication of the buffer sleeve. A soft silicone rubber (Dragon Skin FX- Pro, Smooth-On) was employed as the buffer sleeve. This silicone rubber has a tensile strength of 2.0 MPa, which is considerably larger than the maximum stress calculated from the simulation and hence will not rupture in the measurements. At the first step, a cylindrical silicone rubber tube was fabricated with a molding process, serving as the soft buffer sleeve outside the glass tube that will be detailed in the next paragraph. As shown in Fig. 4(a), a syringe needle (16 Gauge) with an inner diameter of 1.2 mm was filled with uncured, liquid silicon rubber to define the outer surface of the silicone cylinder. One benefit of using a syringe needle is the available and easy assembly of a syringe body with the needle, where the syringe body facilitates the pressure application on the liquid silicone during the molding (Fig. 4(d)). A steel wire with a diameter of 0.25 mm was held inside the needle syringe to define a concentric thorough hole in the molded rubber cylinder. The purpose of the hole was to house the optical fiber and glass tube later in the fabrication. We designed several small acrylic pieces (Parts *i*, *ii*, and *iii* in Fig. 4 (a-c)) to align and support the metal wire and the syringe needle, in order to ensure they were concentric. These acrylic pieces allowed the flow of uncured, liquid silicone rubber into the mold when pressed and avoid air trapping inside the mold. These custom pieces were fabricated by cutting commercial acrylic pieces (8560K172, McMaster-Carr) with a thickness of 1/16" each with a laser cutter (PLS6.150D, Universal Laser), followed by the assembly shown in Fig. 4(c).

After the injection of the silicone rubber into the mold and cure of the rubber, the molded rubber was pulled out from the mold (Fig. 4(e)) and was cut by a knife blade into 3-mm-long sections (Fig. 4(f)).

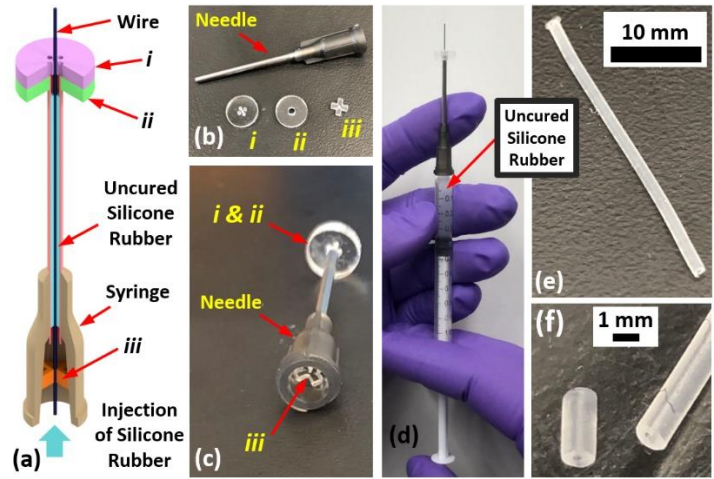

Fig. 4. Phase I of the fabrication process: fabrication of the buffer sleeve. a) Schematic of the molding setup. b) Acrylic pieces and a syringe needle. c) Assembled pieces with a needle. d) Injection of the silicone rubber into the mold. e) Cured silicone rubber tube pulled out from the mold. f) A 3-mm-long silicone rubber tube cut from the long one.

Phase II of the fabrication was to make the reflective cap, which was composed of a glass tube and a reflective copper disk fixed at one of the tube ends. The cap fabrication was carried out in three steps, as shown in Fig. 5(a). First, a copper disk with a diameter of 0.6 mm and thickness of 0.05 mm was cut by a syringe needle tip from a copper sheet. The flat end face of a commercial glass tube (1068150026, Molex), which has outer and inner diameters of 360 μm and 250 μm, respectively, was pushed against the copper disk (Fig. 5(a) Step 1), followed by applying super glue around the tube-disk contact surface (Fig. 5(a) Step 2). The other end of the tube was cut using a capillary column cutter (Capillary GC Column Cutter, Shortix) to obtain a 1.5-mm-long cap (Fig. 5(a) Step 3) that allowed an optical fiber to slide in.

The final phase of the fabrication was the assembly of the buffer sleeve with the reflective end, as shown in Fig. 5(b). First, we inserted a single-mode optical fiber (SMF-28, Corning) tip into the 3-mm-long silicone rubber tube, which was fabricated in Phase I. The fiber has an outer buffer layer made with acrylate, middle glass cladding, and center glass core, with the diameters of 250 μm, 125 μm, and 8.2 μm, respectively. The fiber can freely slide inside the silicon rubber tube and in the glass tube. We slid the tube along the fiber far away from the fiber tip, so it was not shown in Fig. 5(b) until it was slid back to the tip in Step 3. The fiber tip was then inserted into the glass tube (Fig. 5(b) Step 1). The fiber-disk distance was adjusted and simultaneously measured by using the method described in Section 2 and results in Fig. 3(c), until the desired value of $\delta \sim 300$ μm was achieved. As discussed at the end of Section 2, we intentionally chose an initial $\delta = 300$ μm in the fabrication, resulting in an expected $\delta$ range of 100~300 μm when the food shrinks by 20%, to prove the concept of the sensor while retaining a high sensitivity and resolution. Note that $\delta$ is supposed to decrease after the sensor is mounted in foods that shrink in the drying process. A bead of liquid silicone glue (Sil-Poxy, Smooth-On), which was different from the silicone rubber used before, was gently applied to the part of the optical fiber that was outside the glass tube (Fig. 5(b) Step 2). Finally, the silicone rubber tube was slid down along the fiber to fully cover the glass tube, distributing the silicone glue and fixing the sleeve inner sidewall with the tube and fiber (Fig. 5(b) Step 3). The cure of the glue took 12 minutes at the room temperature, completing the sensor fabrication process. A typical fabricated optical fiber strain sensor is shown in Fig. 5(c). If a larger measurement range of food strains is needed, a shorter sleeve can be readily cut to enable a higher buffering ratio, without any changes of the other components or the fabrication procedure.

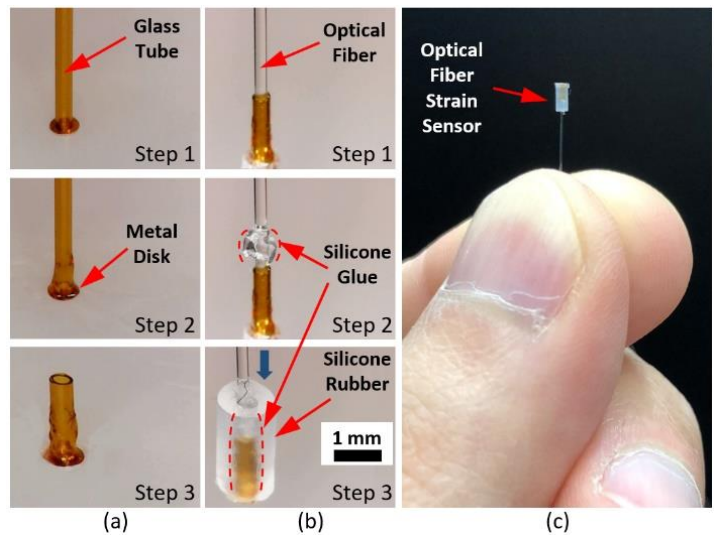

Fig. 5. Phases a) II and b) III of the sensor fabrication process. c) Fabricated optical fiber strain sensor.

## 4- Calibration Setup, Results, and Sensor Performance

We characterized our sensor performance by obtaining the dependence of the sensor readout on the strain of the sample in which the sensor was embedded. The calibration setup, shown in Fig. 6(a), allows a well-controlled strain to be applied to the sensor. A transparent silicone rubber cylinder was molded to mimic a food sample, as shown in Fig. 6(c). In the rubber cylinder, there were a central through hole and four off-axis through holes that were equally spaced from the center and along 4 orthogonal directions. The purpose of the center through hole was to house the sensor, and that of the other four holes was to prevent the bucking of the cylinder under compression. Two translational stages controlled by servomotors (Z825B, Thorlabs) were used to compress the two ends of the silicone cylinder and to control its strain, as shown in Fig. 6(b) and(d). A camera was used to monitor the sensor during the embedment and the calibration processes, thanks for the transparency of the cylinder. The sensor embedment process is depicted by the schematics in Fig. 6(e) and the corresponding photos Fig. 6(f). First, a steel tube with inner and outer diameters of 1.6 mm and 2.1 mm, respectively, was pushed through the central hole from the right, until it reached the other end face of the cylinder. Since the inner diameter of steel tube was larger than the sensor outer diameter, the sensor was readily inserted into the steel tube from the left until it reached the middle of the cylinder. The steel tube was pulled out from the right, leaving the sensor embedded in the middle of the cylinder. With a sensor embedded in a food-mimicking cylinder, as shown in Fig. 7(a), the sensor deforms together with the cylinder. When the compressive strain of the cylinder was 0%, 10%, and 20%, the fiber-disk distance $\delta$ became smaller in turn, as can be seen in Fig. 7(b).

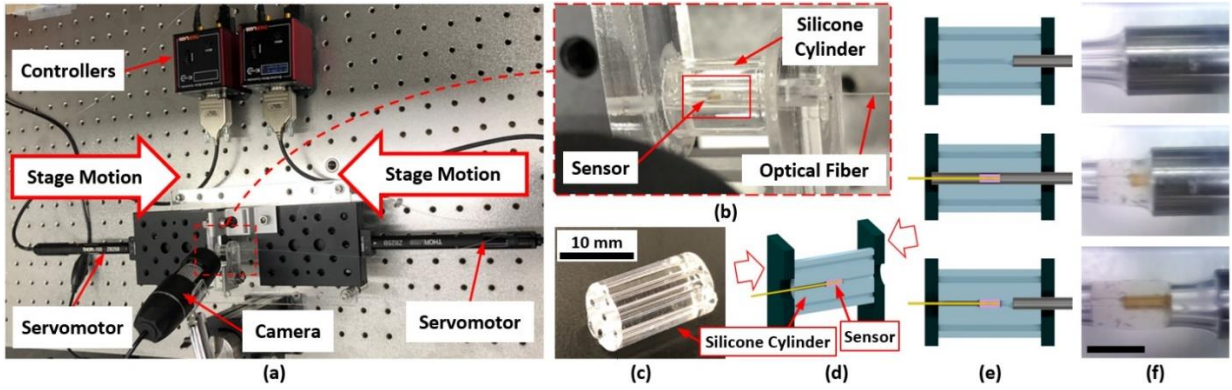

Fig. 6. a) Calibration setup and b) a close-up. c) Transparent silicone rubber cylinder to mimic the food. d) Schematic of the cylinder between stages with the sensor inside. e) Schematics and f) photos to depict the steps of inserting the sensor inside the cylinder (top to bottom). Scale bar in (f) = 2 mm.

We used this setup to calibrate 12 different sensors by obtaining the dependence of the cylinder strain on the collected optical power. We assumed the compressive strain in the axial direction was uniform in the silicone cylinder, because of the homogeneous and isotropic mechanical properties of the silicone, the rotationally symmetric geometry of the cylinder, and the axial compressive loading on two large flat end faces of the cylinder. Therefore, we considered the cylinder strain at the sensor location to be equal to the average axial compressive strain in the whole cylinder, the latter of which was obtained by the ratio of the cylinder length change to the initial length. By using this way to obtain the cylinder strain, the statistical calibration results of 12 measurements from 4 different sensors, with 3 measurements each sensor, are plotted in Fig. 7(c). It can be seen that the sensors were able to measure the cylinder compressive strain up to 20%, confirming the validity of our sensor design. As explained in Section 2, the strain rose exponentially and monotonically with the collected optical power in the experimental measurements. The small standard deviations throughout the 0~20% strain range indicate that the molding-based fabrication renders a good repeatability and a good controllability of the sensor performance, promising a high potential for automatic and batch fabrication processes. Next, we would like to find a theoretical model to describe the experimentally measured calibration data. By considering Eq. 3, the strain on the outer surface of the sensor buffer sleeve, which is equal to the silicone cylinder strain at the sensor location, can be expressed as

$$\varepsilon = \frac{\Delta L}{L_0} = \frac{\Delta L}{\Delta \delta}\frac{\Delta \delta}{L_0} = b_r \frac{\delta - \delta_0}{L_0}, \tag{5}$$

where $\delta_0$ and $L_0$ are the initial fiber-disk distance and initial buffer sleeve length, respectively, before loading. $\delta_0$ and $L_0$ were set to be ~300 μm and 3 mm, respectively, in the fabrication process. For a fabricated sensor, $b_r$ is fixed, so $\varepsilon$ changes linearly with $\delta$. With Eqs. 4 and 5 in consideration, the dependence of the change of measured photodetector voltage, which is proportional to the collected optical power, on the cylinder strain can be described as

$$\Delta V = -A \left( e^{\frac{-2}{1+(C\varepsilon+D)^2}} - e^{\frac{-2}{1+D^2}} \right), \tag{6}$$

where $\Delta V$ is the change of photodetector readout voltage, $A$ is 11.6 V determined in Section 2, $C = 2\lambda L_0/\pi\omega_0^2 b_r$, and $D = 2\lambda\delta_0/\pi\omega_0^2$. In practice, both $C$ and $D$ are constants to be determined by the fitting to the experimental data. The second term on the right-hand side of Eq. 6 is to allow for the vanishing of voltage change at zero strain. It is noted that compressive strain was experienced both by the silicone cylinder and by the fiber sensor in Eq. 6. We used Eq. 6 to fit the experimental calibration data, with the fitted curve plotted in Fig. 7(c). A fitting can be obtained with a $R^2$ of 0.999, while the fitting parameters $C$ and $D$ are 0.183 and 6.225, respectively. Such a good fitting indicates that our theoretical model of Eq. 6 closely describes and predicts the sensor performance. We note that the data in Fig. 7(c) are the most important results of the calibration. With these data, one can embed the sensor in any food samples and can measure unknown food strains based on the photodetector readout of the sensor.

Before describing the sensor demonstration in real foods, we will use a few commonly used sensor characteristics and the buffering ratio, the last of which is specific to our sensor, to understand the performance of our strain sensor.

**Dynamic range.** The dynamic range describes the range of the desired parameter that can be reliably measured. The dynamic range of our fiber strain sensor is 0~20% compressive strain. According to Fig. 3(c) and Fig. 7(c), the dynamic range corresponding to 50 μm < $\delta$ < 300 μm. Note that we did not realize a strain higher than 20% in the experiment, because the silicone cylinder started to buckle. Therefore, we did not test our sensor under the sample strain higher than 20%, and it is possible that the higher boundary of the dynamic range is >20%.

**Sensitivity.** Sensitivity is another important characteristic of sensors and is defined as the derivative of sensor readout with respect to the input. In our fiber strain sensor, the input and readout are the strain to be measured and photodetector voltage change, respectively. The sensitivity is essentially the slope of the experimentally measured curve in Fig. 7(c). The dependence of sensitivity on the strain is plotted in Fig. 7(d). The sensitivity depends on the cylinder strain. The sensor is more sensitive as the cylinder deforms more, which is expected because $\delta$ is smaller and the working condition of the sensor is closer to the left-hand side of the curve in Fig. 3(c). Such a strain-dependence of the sensitivity is particularly beneficial in food drying, because the sensor provides more sensitive strain measurements at the later stage of drying, in which stage the strain information is more important to make a proper decision on stopping the process.

**Resolution.** Resolution of the sensor can be calculated as $\Delta\varepsilon = V_n/Sensitivity$, where $V_n$ is the noise in voltage in the photodetector readout of the sensor. In our fiber strain sensor, the sensor noise could be caused by three sources: the light source noise, photodetector noise, and mechanical drifts between or instability of different sensor components. All the three the noise sources are manifested as the fluctuations of photodetector readout, when the embedded fiber sensor is deformed by the silicone cylinder and held in place. In principle, such fluctuations depend monotonically on the sensor signal which in turn depends on the silicone cylinder strain. In practice, due to the difficulty to quantify such fluctuations at every possible cylinder strain, we experimentally measured such fluctuations to be 0.010 V, 0.014 V, and 0.020 V at the cylinder strains of 0, 10%, and 20%, respectively. Therefore, we overestimate the photodetector noise over the whole cylinder strain range of 0~20% to be 0.020 V. In particular, in the $\delta$ range of 900~1000 μm in Fig. 3(c), where light reflection from the mirror can be neglected and the last noise source vanished, we find the photodetector voltage fluctuation to be ~0.002 V, significantly smaller than 0.020 V. This fact validates the noise estimation of 0.020 V because only two of the noise sources, namely light source and photodetector noise, existed and the sensor signal was smaller in this range than the that in the sensor dynamic range. Since sensitivity is not constant, the resolution depends on the strain. The highest and lowest resolution of our sensor are ~0.08% and 0.72% in absolute value of normal strain, occurring at the strains of 20% and 0, respectively.

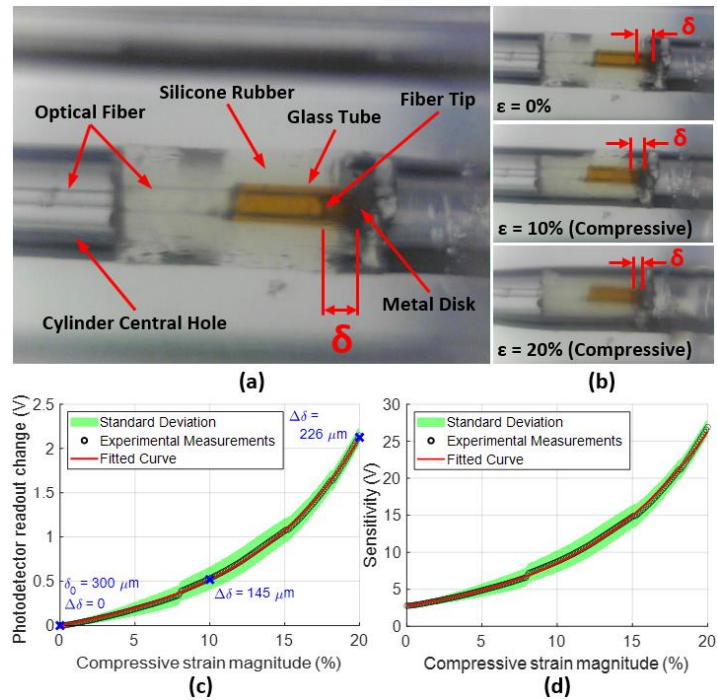

Fig. 7. Calibration results obtained through 12 measurements from 4 different fiber strain sensors. a) Photo of a sensor embedded inside the food-mimicking silicone cylinder. b) Photos of sensor responses to different cylinder strains. c) Cylinder strain versus the photodetector readout change. Three points on the curve were labeled by blue crosses, corresponding to the three photos in (b). The three labeled $\Delta\delta$ values

are calculated by combining the measured photodetector readout changes and the results in Fig. 3(c). d) Sensitivity of the sensor, which is the derivate of photodetector readout with respect to strain, versus the strain. Since the strain is dimensionless, the unit of the sensitivity is the same as that of the photodetector output (V).

**Buffering ratio $b_r$.** Using the dependence of photodetector voltage change on $\delta$ plotted in Fig. 3(c), we can convert the voltage in Fig. 7(c) to $\delta$, which is plotted as the red solid curve in Fig. 8(a), and in turn find $\Delta\delta$ by $\delta - \delta_0$. Since $\Delta L$ can be readily calculated by $\varepsilon \cdot L_0$, which is plotted as the green dashed line in Fig. 8(a), we are able to determine $b_r$, by $\Delta L/\Delta\delta$ following Eq. 1, at any strain from the measured calibration data. The obtained $b_r$ is plotted as the red curve in Fig. 8(b). It is noted that the experimental buffering ratio is always less than 3, which is expected from the FEM simulation. This discrepancy might result from the imperfect binding between the buffer sleeve and the fiber or the glass tube, which most likely occurred at the fiber entrance of the glass tube. The geometry of the entrance of glass tube imposed a challenge on the conformability of the buffer sleeve step coverage, especially after the fiber was moved into and out of the tube caused by the sensor deformation. Such a challenge might lead to the delamination of the buffer sleeve from the local fiber or glass tube, while this delaminated portion of buffer sleeve did not function to scale down the deformation on the fiber-disk distance, leading to a smaller $b_r$. In general, a partial binding with the buffer sleeve anywhere along the fiber or glass tube could lower the buffering ratio which was observed in the experiments.

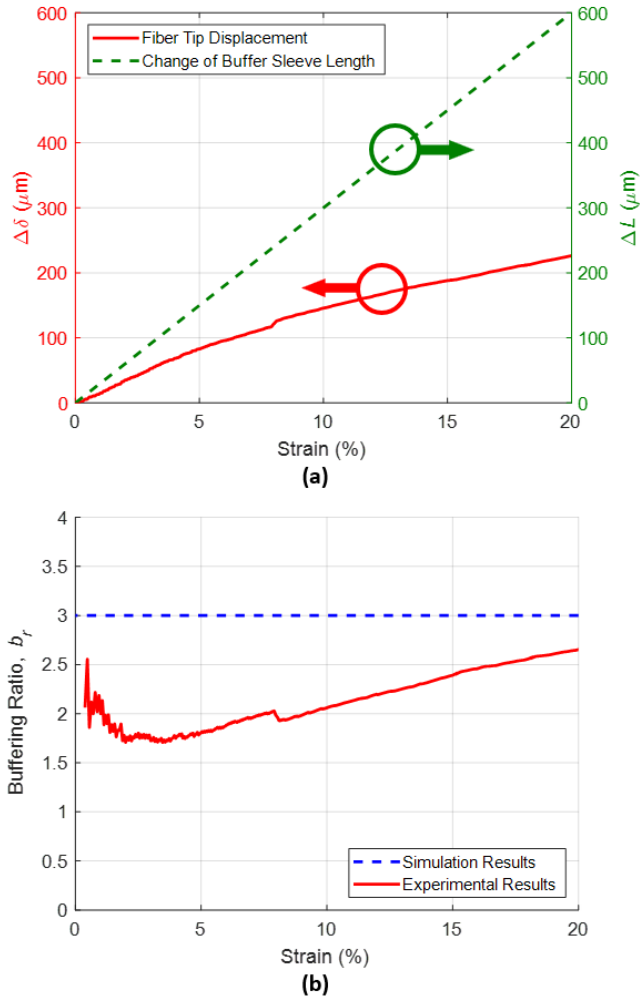

Fig. 8. Evaluation of the buffering ratio from the experimental calibration results. a) Dependence of the length change on the outer sidewall of buffer sleeve ($\Delta L$, green dashed line) and the fiber tip displacement ($\Delta \delta$, red solid curve) on the cylinder strain. b) Buffering ratio ($b_r = \Delta L/\Delta \delta$) at different cylinder strains.

## 5- Sensor Demonstration in the Drying of Fresh Banana Slices

With the understanding of sensor performance obtained from the calibration, we further demonstrated the application of the optical fiber strain sensors in foods by continuous measurements of the shrinkage below the surface of fresh banana slices during drying. We chose the fresh banana as the food sample, because it is a common fruit readily available across the world. It is challenging to measure under-surface strains of banana slices with conventional methods, due to the soft texture, high moisture content, non-uniform and large deformation in drying, and large mechanical stiffness change before and after drying. Therefore, fresh banana slices make a good example of foods to demonstrate the capability of our optical fiber strain sensors.

The experimental setup is presented in Fig. 9(a). Cavendish bananas with a Brix value between 20% and 25% were purchased from local food markets and cut into slices, each with a thickness of ~7 mm and a diameter of ~30 mm. The banana slice was placed inside an oven with an electrical heating element at the bottom and a camera above for imaging. The drying of the banana slices was conducted for four hours at 80 °C. There was no fan inside the oven and the air currents were a result of gravitational convection. A thermocouple was placed next to the banana slice inside the oven and was connected to the electrical power source for the feedback control of temperature, as shown in Fig. 9(b). The wet basis moisture contents of the banana slices were 80±2% before and 40±5% after the drying. As depicted in Fig. 9(c), the fiber strain sensor was inserted into a hole created from the banana slice sidewall along the radial direction, about 2 mm below the top surface. Thus, the fiber sensor provided the measurements of the compressive normal strains in the radial direction of the banana slice at the sensor location. We note that we chose the radial direction to demonstrate the sensor measurements without specific reasons. The fiber sensor can be inserted in any desired direction into the food sample and will be able to provide the normal strain measurements along that direction.

Due to the banana softness and wet surfaces, which are challenges of any strain sensor implementation, we would provide some details of our fiber sensor implementation to ensure a good contact and grip between the banana and sensor. The hole in banana was created with a metal tube with an outer diameter of 1 mm which is slightly smaller than sensor diameter (1.2 mm), resulting in the banana gently squeezing on the outer surface of the embedded sensor. The tight fit helped to achieve a uniform contact between the banana and the cylindrical outer surface of the sensor and hence a reliable sensor attachment inside the banana. To implement the sensor in a banana slice, we held the fiber end of the buffer sleeve and push the fiber sensor into the hole by hand, without using any starch or glue between the banana and sensor. There was enough amount of starch in the fresh banana that provided gradually enhanced grip on the fiber sensor during the drying. That said, the grip might not be the most reliable in the beginning of the drying process, where the banana deformation was small and less important, but the grip was better and sensor readout more reliable toward the drying end, where the deformation was large and the measurements more crucial.

The fiber sensor tests were repeated with a total of 4 sensors and 14 banana slices with the same thickness cut from the same banana finger. A different banana slice was used in each test with one fiber sensor embedded at the same radius and depth in the slice. Each sensor was reused for 3~4 tests until the sensor surface got contaminated and performance degraded.  To reuse the sensors, we were able to remove the sensors by soaking the dried banana slices in water and softening them. We were aware of the non-uniform distribution of the banana mechanical properties along the azimuthal direction, which was difficult to observe before the sensor insertion. Therefore, we chose a random radial direction of each banana slice to insert the sensor along and obtained an average strain from all the sensor measurements in 14 slices, with a purpose of removing the azimuthal dependence. The final expected results are the time-dependent strain continuously measured by the fiber sensors at a fixed radial position on all 14 banana slices over the whole drying process.

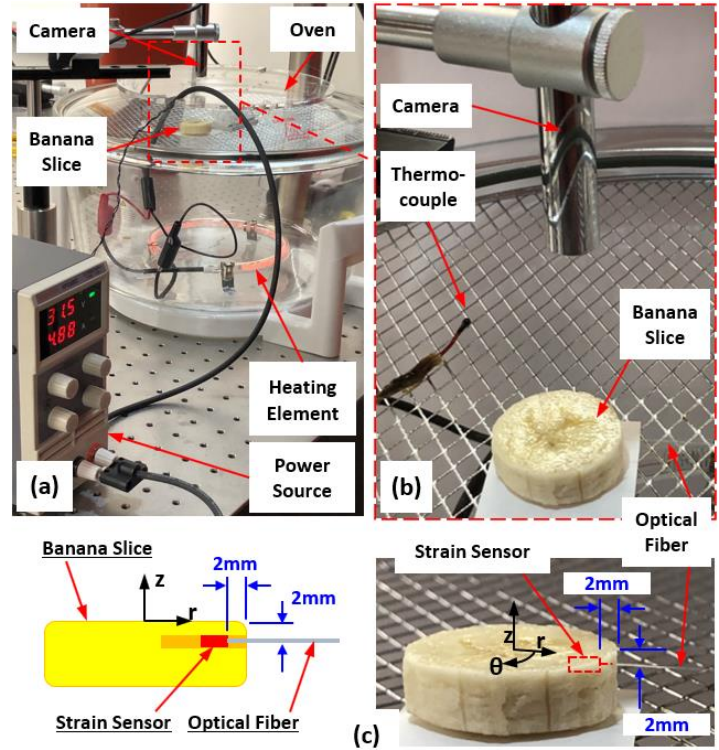

Fig. 9. Sensor demonstration in the drying of fresh banana slices. a) Drying setup. b) Close-up. c) (top) Schematic and (bottom) photo of a banana slice with the sensor embedded.

The computer vision (CV) method was used to analyze the camera images as an alternative measurement approach of strains on the top surfaces of banana slices, serving as a comparison and validation for our fiber strain sensor measurements. 4 different banana slices, which were cut from the same banana finger as the ones used for fiber sensor tests, without any fiber sensors embedded. To facilitate the CV measurements, 32 ink dots were marked, before drying, along eight radii on the top surface of each banana slice, with an angle of 45º between each adjacent radii, as shown in Fig. 10(a). As the banana slice was dried and shrank, these dots moved accordingly, as shown in Fig. 10(b), were continuously tracked by the camera images during drying, and their position changes were analyzed by the image-processing after drying to calculate the strain between each two adjacent dots. Along each radius, we marked 4 dots that were ~3 mm apart from each other, with the outermost dots ~2 mm away from the circular edge. Note that the outermost two ink dots in each radial direction were co-located with the two ends, respectively, of the fiber sensors that were embedded in the abovementioned sensor tests. To remove the azimuthal dependency of the strains in each slice, we averaged the strains measured by all the outermost dot pairs in 8 radial directions to obtain the CV-measured strain in this slice, for the purpose of comparison with the fiber sensor measurements. Therefore, the final CV measurements of the time-dependent strain at the fiber sensor radial location were obtained by averaging those obtained from all 4 banana slices at any given time in the drying process.

It is worth noting that the CV measurements were limited on the banana slice surface, while the fiber strain sensors measured strains 2 mm below the surface, so there should not be a perfect match between the measurements of CV and the fiber sensor due to the location difference. Since there is no existing method to measure strains below the banana surface, which is why our fiber strain sensor is unique and beneficial, CV measurements still provided a reference to validate the developed fiber strain sensors.

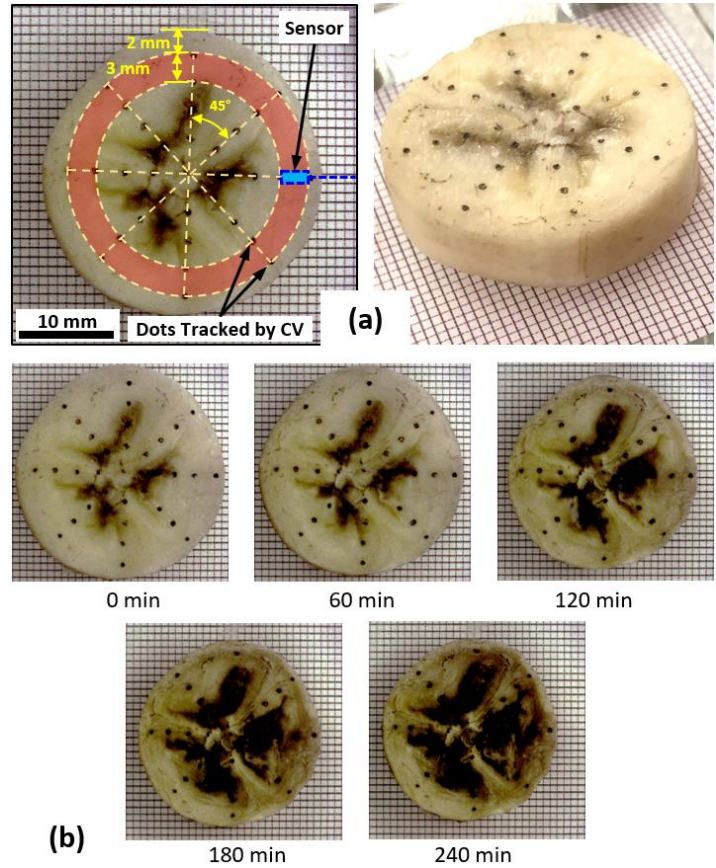

Fig. 10. Computer vision (CV) based surface strain measurements. a) Sample preparation with ink dots marked on the banana slice top surface. In the left figure, the dots were distributed along eight different radial directions (dashed straight lines). The red ring indicates the area at which the CV-based measurements were carried out. The inner and outer radii of the red ring coincide with 8 pairs of 2 dots that were used for the strain measurement. 2 dots of each pair were at the same radial location of the ends of the embedded fiber strain sensor that is shown with blue. b) Photos of the banana slice taken at different time in the drying process. All the photos share the same magnification and viewing angle.

The change of photodetector readout of the fiber strain sensors during the whole drying process is shown in Fig. 11(a). As the banana shrank, the measured voltage increased, indicating a reduction in the fiber-disk distance $\delta$. Using the model described in Eq. 6, one can convert the photodetector readout change at any time point in Fig. 11(a) into the banana strain measured by the fiber sensor. The resulting fiber sensor strain measurements, as well as the CV measurements, of the banana slice are shown in Fig. 11(b). There is a good agreement between these two measurements, and the discrepancy in between reveals interesting information of the banana slice drying. First, the CV measured strains are higher throughout the drying, with a strain difference of 2~5% except in the first 10 minutes. The larger CV measurements are expected because the air-drying process removes the moisture more efficiently from the surface than internal, resulting in the surface deformation always larger than internal deformation. Such a difference between surface and internal behavior is also referred to as the shell-hardening effects [8]. Time dependence-wise, the slope of the CV measurements was highest in the beginning, gradually reduced in the first 50 minutes, and stays less changed in the rest of the drying process. This indicates that the surface drying and hence the surface deformation were fastest in the beginning due to the high initial moisture content, slowed down due to the reduced moisture contents and enhanced difficulty of moisture removal, and finally became stable when most of the removed moisture came from the food volume through the surface. By comparison, the fiber sensor measured strain started very slow before 40 minutes, rose fast from 40 to 90 minutes to catch up the CV measurements, and stays stable and almost always parallel to the CV measurements after ~90 minutes. This result confirms that the air-drying started from the surface and that the moisture removal from below the food surface, as well as the resulting deformation, occurred with a time delay. The comparison between the time dependence of the two measurements indicates that the moisture removal from the banana surface was dominant in the first 40~50 minutes of drying, after which time the dominant drying effect was to remove the moisture below the surface. The effective moisture removal at 2 mm below the surface, where the fiber sensor was embedded, ended at ~90 minutes. We hypothesize that the effective removal of the moisture after 90 minutes occurred deeper in the food matrix, and the study of the depth dependence of the strain is underway. The above discussion implies that the fiber optical strain sensor, by providing previous unavailable information in food study, has a high potential to provide new understandings of the food drying process.

We note that the strain of the banana slice in Fig. 11(b) reached ~20%, measured by both the CV and fiber sensor, proving that our fiber sensor design goal of measuring 20% food strains has been achieved. As discussed in Section 2, the design parameters, including the geometrical and material properties of the components, can be readily changed to obtain a higher buffering ratio, further extending the measurable strain range. For example, a higher buffering ratio can be obtained by increasing the thickness, decreasing the length of the buffer sleeve, or choosing softer silicone rubbers to fabricate the buffer sleeve.

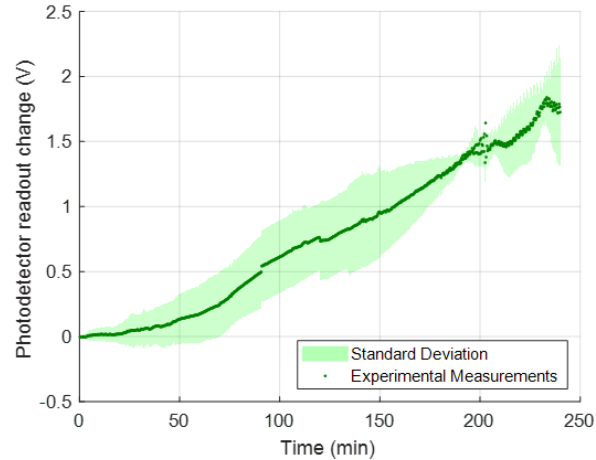

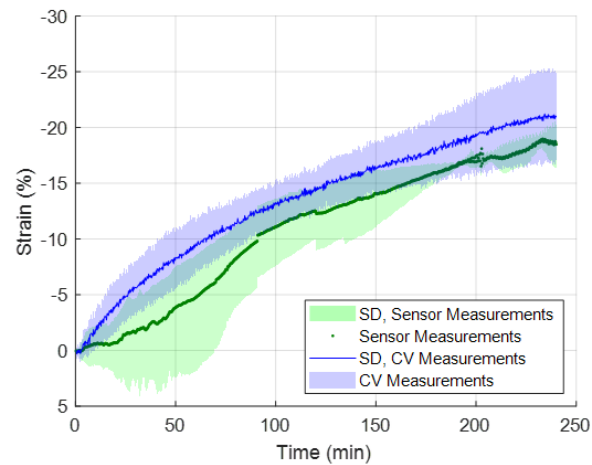

Fig. 11. Optical fiber strain sensor and CV measurements. a) Experimentally measured signal from 14 tests obtained from 4 optical fiber strain sensors. b) Calculated strain from optical fiber strain sensor signal by using calibration data and CV measurements. CV measurements were done on 4 banana slices where 8 strains were measured from each slice (32 strain measurements in total). SD = Standard Deviation; CV = Computer Vision.

## 6- Conclusion

We present an optical fiber strain sensor for the measurements of food deformation during the whole drying process. To the best of our knowledge, our work demonstrates the first strain sensor that enables continuous deformation measurements below the food surface and the first one that is mechanically flexible enough to be embedded in soft food samples, such as fresh banana slices, with minimal changes to the native food deformation. The small sensor footprint allows the measurements of deformation at a mm-scale location in a food sample, as well as those inside small food samples such as grains, while the optical fiber-based light readout bestows the sensor the compatibility with harsh environments such as those with strong electromagnetic interferences. One of the key design features of our fiber sensor is the soft buffer sleeve on the sensor outer surface, enabling a large measurable deformation range and minimizing the sensor influence on the native food deformation. The fiber sensor can be embedded under a constant pressure from food samples that provides a good attachment. The buffer sleeve was fabricated by a molding process, resulting in highly repeatable sensor performances. The rest of the fabrication and assembly processes were straightforward without requiring any specialty facilities. To obtain a systematic understanding of the sensor performance, we have experimentally determined the sensor characteristic attributes, including sensor responses to external strains, dynamics range, sensitivity, and resolution. Moreover, we applied the fiber strain sensors to continuously measure the deformation below the surfaces of fresh banana slices during a 4-hour-long drying process, in order to demonstrate the capability and faithfulness of the sensor for challenging food samples. The banana slices were wet and soft in the beginning and were dry with glassy textures at the end, with a wet basis moisture content of around 80% and 40%, respectively. Computer vision (CV), in which imaging processing of the photos of the banana top surfaces were carried out, was employed to measure the strains on the surfaces of the banana slices, serving as a reference for the fiber strain sensor measurements. We obtained a good agreement in the results by the optical fiber strain sensor and CV, which proves the faithfulness of the fiber strain sensor measurements. The discrepancies between the results of the two methods reveal the differences between the surface and below-the-surface strains. Specifically, both the spatial and time dependences of the measured strains showed characteristics of shell-hardening effects. The sensor tests in the banana slice drying demonstrate the fiber strain sensor capability of continuous, in-situ, and below-the-surface strain measurements and indicate that such an unprecedented sensor has a high potential to contribute to improving both fundamental understandings and process monitoring of food drying processes.

## Acknowledgments

This study was partially supported by the US Department of Energy (# DE-FOA-0001980), the Center for Advanced Research in Drying (CARD), and the Massachusetts Clean Energy Center (MassCEC). CARD is a US National Science Foundation Industry University Cooperative Research Center. CARD is located at Worcester Polytechnic Institute, with the University of Illinois at Urbana-Champaign as the co-site.